# Financial Market Trend Forecasting and Performance Analysis Using LSTM

Jonghyeon Min

# Abstract

Financial market trend forecasting method is emerging as a hot topic in financial markets today. Many challenges still currently remain, and various researches related thereto have been actively conducted. Especially, recent research of neural network based financial market trend prediction has attracted much attention. However, previous researches do not deal with financial market forecasting method based on LSTM which has good performance in time series data. There is also a lack of comparative analysis in the performance of neural network based prediction techniques and traditional prediction techniques. In this paper, we propose a financial market trend forecasting method using LSTM and analyze the performance with existing financial market trend forecasting methods through experiments. This method prepares the input data set through the data preprocessing process so as to reflect all the fundamental data, technical data and qualitative data used in the financial data analysis, and makes comprehensive financial market analysis through LSTM.

In this paper, we experiment and compare performances of existing financial market trend forecasting models, and performance according to the financial market environment. In addition, we implement the proposed method using open sources and platform, and forecast financial market trends using various financial data indicators.

Table of Contents

# 1. Introduction

Forecasting trends in financial markets by analyzing financial market data is still challenging in many areas. This is because many factors, such as political reasons, economic conditions, and intuition of investors, affect financial markets. Equity and futures traders have sought to help with various types of intelligent systems to make investment decisions, but so far the success of existing systems has been very limited. Because financial market trends are non-linear, uncertain, and non-stationary, even financial data professionals find it difficult and complex to accurately predict financial market trends using financial market data. Nonetheless, in today's financial markets, investment banks and hedge funds such as Goldman Sachs, Quantum Funds and Renaissance Technologies are in the process of developing intelligent systems and quantitative algorithms that help to overcome the psychological bias. Also, recently, system development using machine learning techniques has been actively carried out due to the development of machine learning and computing ability and data explosion.

In the meantime, there have been various attempts to predict financial markets from traditional time series data approaches to artificial intelligence techniques including the ARCH-GARCH model, the ANN, and the method of using evolutionary algorithmic. Among them, ANN is known to be more effective than other models in forecasting financial market trend. Previous studies have shown that models using ANN perform better than statistical regression models and discriminant analysis.

In general, there are two methods of making financial market forecasts using ANN. The first methodology is to consider financial market fluctuations as a time series and to forecast future trends using historical data. In this methodology, ANN is used as predictors. However, these forecasting models are limited in performance due to numerous factors affecting noise and stock prices. As a result, none of the existing forecasting models using this methodology has shown satisfied performance. The second methodology is a methodology for forecasting and analyzing trends in financial markets, taking into account qualitative factors that affect financial markets or trends, such as political influence. This methodology provides a robustness to deal with imprecision, uncertainty, and partial truth and also cost-effective solution as qualitative and technical indicators are combined. It shows a meaningful performance compared to the first methodology.

The LSTM is a neural network structure that improves the RNN in order to solve the long-term dependency problem in RNN. Since the short and long term storage networks show great performance in modeling time series data, various problem areas using time-series data, such as natural language processing or image processing, perform better than other network architectures and are used in many areas.

Financial market data is suitable for the structures of the LSTM because of the characteristics of time-series data suitable for the LSTM. However, there is a lack of research on the financial market trends forecasting methods using the LSTM in the existing studies, and there is also a lack of research comparing, analyzing, and evaluating the performance of the method from various viewpoints.

In this paper, we propose several financial market forecasting methods using the LSTM and verify the performance of the proposed methods through various experiments such as performance comparison experiments with existing models from various viewpoints and performance comparison depending on the financial market condition. In addition, the proposed methods reflect all data of fundamental analysis, technical analysis, and sentiment analysis.

## 2. Theoretical Backgrounds

Regarding the research in this paper, it is necessary to know background knowledge of the Recurrent Neural Networks, the Long-Short Term Memory Networks, Financial Data Analysis Methodology, and Studies on Financial Market through Data Analysis.

### 2.1 Recurrent Neural Networks

The Recurrent Neural Network (RNN) is a neural network structure that shows great performance in various tasks with the time series data. The purpose of the RNN is to process sequential information. In the conventional neural network structure, all inputs and outputs are independent of each other. However, many cases of real data have a time series data structure that changes with time and has a context relation between data. The RNN is the appropriate neural network structure having this type of structures. If financial market data were to be used to predict trends in the financial market, it would be better to use the dependent relationships between the data to reflect the financial market data in the results, rather than assuming separate financial market data. Since the RNN has a recurrent structure literally, the result of the previous element affects the result of the next element. That is, the RNN has memory information on the calculated result so far, and the information up to now affects the next result. The basic structure of the RNN is shown in [Figure 1].

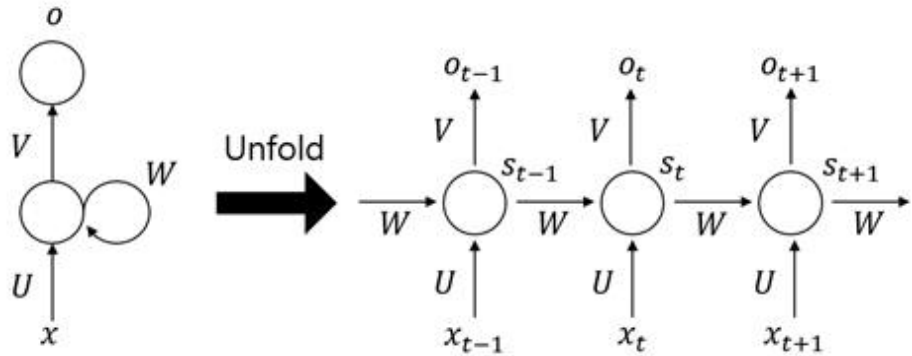

[Figure 1] The basic structure of the RNN

[Figure 1] shows that the recursive connection of the RNN is unfolded. $x_t$ is the input at time step t. $s_t$ is the hidden state value at time step t. This part acts like the memory of the network. It is calculated as <Equation 1> by the hidden state value ($s_{t-1}$) at the previous time step t-1 and the input value ($x_t$) at the current time step t.

<Equation 1> Computation of the RNN Hidden State

$$s_t = f(Ux_t + Ws_{t-1})$$

$f$ is an activation function, usually [tanh] or [RLU] are used, which is a nonlinear function. Usually, it is set that $s_{t-1}$ for calculating the first hidden state is zero. $o_t$ is calculated using the value of $s_t$ as the output at the time step t. $U$ and $W$ indicate respectively weights.

There is a problem that the RNN can not solve the Long Term Dependency problem which makes it difficult to connect the context of the information as the order difference becomes larger.

### 2.2 Long-Short Term Memory Networks

It is difficult to say that the LSTM has a structure that is inherently different from that of the RNN. However, as shown in [Figure 2], we use a different structure from the RNN in calculating the hidden state. The LSTM is a special type of the RNN designed to solve the Long-Term Dependency problem of the RNN.

In the LSTM, a structure called a memory cell is used instead of a neuron of the RNN, and it can be considered as a black box that receives an input of a previous state $h_{t-1}$ and a current input $x_t$ as an input value. Inside the memory cell, it decides how much to retain and how much to remove the previous memory value, and calculates the value to be stored in the current memory based on the current state and the input value of the memory cell.

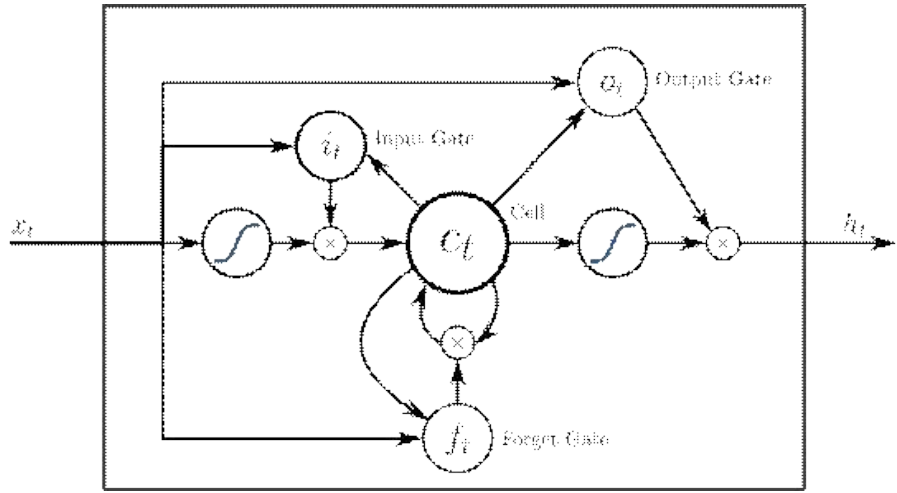

[Figure 2] The Basic Structure of the Long-Short Term
Memory Networks

The core of the LSTM is cell state. The LSTM can add or erase information to the cell state. This is controlled by a structure called a gate, which consists of multiplying the neural network system layers and the elements, as shown in <Equation 2>. The first step in the LSTM is to decide which information to discard from the cell state. This is determined in the "forget gate". Forget gate vector ($f_t$) takes the $h_{t-1}$ and $x_t$ as inputs, and is calculated using weights ($W$, $U$) and bias ($b$), and then takes the activation function ($\sigma_g$). The value comes out between 0 and 1, and the closer to zero, you use the less information in the past. Next, decide which values to update in the "input gate". The input gate vector ($i_t$) also takes the $h_{t-1}$ and $x_t$, is calculated using weights and bias, and takes the activation function ($\sigma_g$). The memory cell state ($c_t$) is determined by multiplying the previous memory cell state ($c_{t-1}$) and $f_t$, and adding the value multiplied by $i_t$ and the activation function ($\sigma_c$) to the same input value. Finally, in the "output gate", the final value ($h_t$) is determined through the activation function ($\sigma_h$).

This structure is more effective in dealing with long time series data. Because financial market data used in this paper is long and the extent to which historical information affects future information is more important than other data, it is appropriate to use the LSTM for financial market trend forecasting method.

<Equation 2> LSTM's Memory Cell Calculation Process

$$f_t = \sigma_g(W_f x_t + U_f h_{t-1} + b_f)$$
$$i_t = \sigma_g(W_i x_t + U_i h_{t-1} + b_i)$$
$$o_t = \sigma_g(W_o x_t + U_o h_{t-1} + b_o)$$
$$c_t = f_t \circ c_{t-1} + i_t \circ \sigma_c(W_c x_t + U_c h_{t-1} + b_c)$$
$$h_t = o_t \circ \sigma_h(c_t)$$

**2.3 Financial Data Analysis Methodology**

Financial data analysis methodologies are divided into fundamental analysis, technical analysis, and sentiment analysis. Fundamental analysis refers to the analysis of fundamental factors and intrinsic value.

The fundamental analysis is based on the assumption that there is a gap between market value and intrinsic value, and product price converges to intrinsic value. However, the problem with data in financial statements is that timeliness may lead to inconsistencies between market value and intrinsic value. Also, there is a problem with the difficulty of systemic verification due to the reduction of objectivity in valuation of intrinsic value due to subjective intervention of experts.

The technical analysis refers to a method of analyzing the movement of stock prices using a variety of indicators such as changes in past stock prices and trading volume. The technical analysis is based on an efficient market in which all information is reflected in prices, moves with a certain trend, and stock price movements repeat similar movements. Despite the controversy about predictability, the technical analysis method is widely used in the field as a tool to grasp trends in financial markets.

The sentiment analysis is a method of forecasting financial market trends by analyzing psychological factors such as investors' greed and fear to financial markets. Subjective information is extracted through natural language processing and text analysis, and various emotions on the target are analyzed and used for decision making. However, the sentiment analysis is rarely used independently in financial data analysis. It is more relevant as an analysis to support the fundamental analysis and technical analysis.

In order to reflect the characteristics of all three analytical methods, we propose techniques to forecast financial market trends using all data for fundamental analysis, for technical analysis, and for sentiment analysis.

## 3. Financial Market Trend Forecasting Models

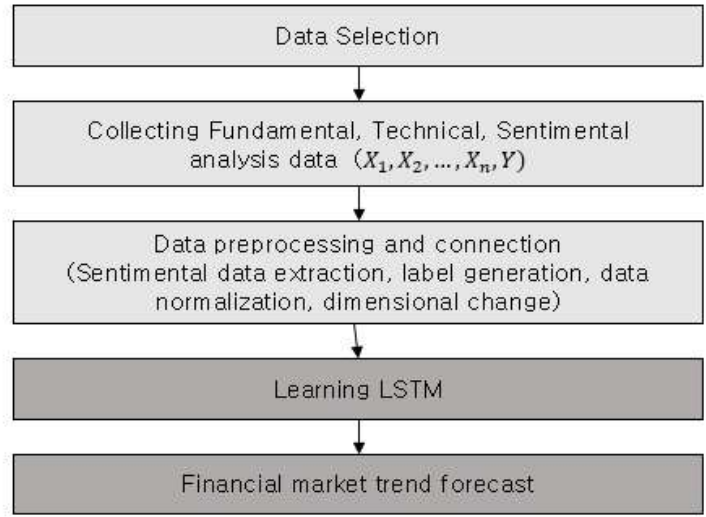

[Figure 3] Financial Market Forecast Model

The financial market forecasting model presented in this paper is divided into two stages as shown in [Figure 3]. The first step is the process of creating a set of data, collecting data for fundamental analysis, data for technical analysis, and data for sentiment analysis. In the fundamental analysis and the technical analysis, various indicator data and quantitative indicator data are used. In the experiment of this paper, several important indicators are selected for convenience of the calculation process, but various other indicators can be used. Although the fundamental analysis has many points to consider besides the indicators, we use indexes that can be quantified for usability of system reflection. After selecting the data to be used, collect data for each. The collected data is subjected to a preprocessing and connection process. Secondary indicators, quantitative indicators, and sentiment data selected for fundamental, technical, and sentiment analysis are connected to each other as input data, as shown in Equation (6). $A_t$ is input data composed of subsidiary indices for fundamental analysis, $F_t$ is input data composed of indices for technical analysis, and $S_t$ is input data composed of sentiment data for qualitative analysis. Each input data increases or decreases the dimension of the data to have the same dimension ($d_I$) using a neural network so as to have an even effect on the prediction results. The final input data $I_t$, which is made up of the same dimension $X_t, Y_t, Z_t$, is used in conjunction. $W_A, W_F, W_S$ represent weight values, and $b_A, b_F, b_S$ represent bias values.

<Equation 6> Input data preprocessing process

$$X_t = W_A A_t + b_A$$
$$Y_t = W_F F_t + b_F$$
$$Z_t = W_S S_t + b_S$$
$$I_t = [X_t\ Y_t\ Z_t]$$
$$W_A \in R^{d_I \times d_A},\ W_F \in R^{d_I \times d_F},\ W_S \in R^{d_I \times d_S}$$

In addition, the process includes extracting sentiment data using an article, generating label data, and normalizing the data. Details of the process are covered in Chapter 4.1 Data Sets.

The second step is to predict the financial market trend by inputting the input data as shown in [Figure 4] as the input value of the LSTM. The goal of the neural network is to predict the adjusted price of the next time step t-1.

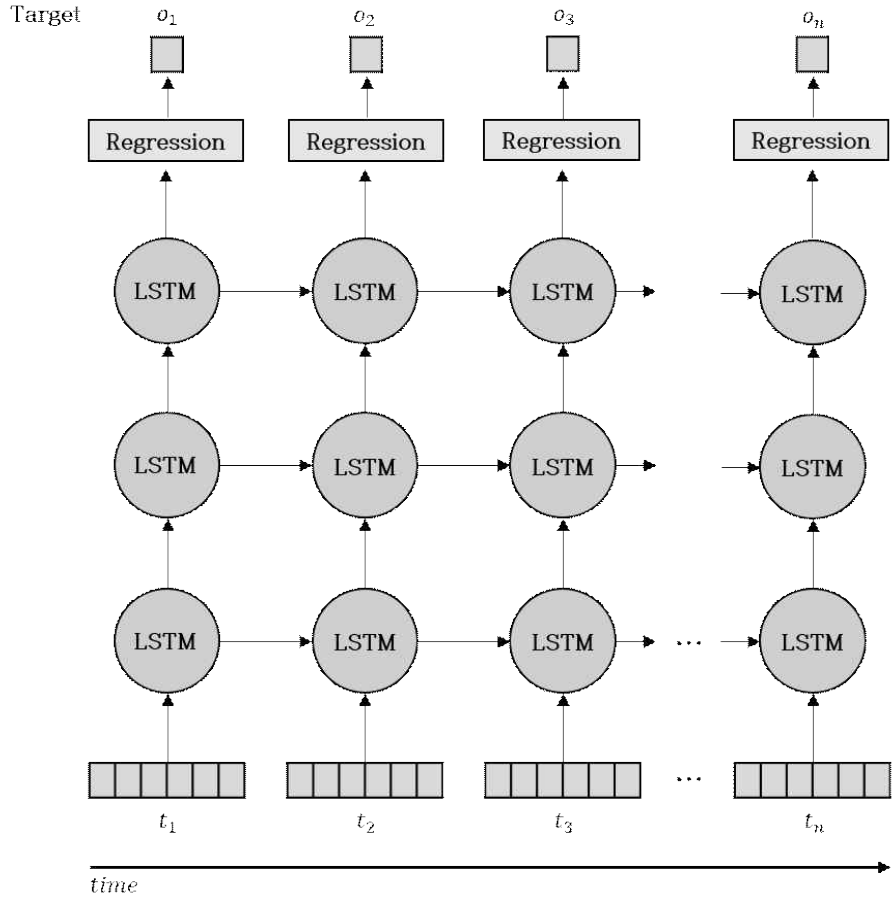

[Figure 4] LSTM in financial market forecasting

LSTM learns by using the input data, and then, in the last step, it produces a value ($o_n$) that predicts the financial market trend through a linear regression process. The output value ($y^{(i)}_{pred}$) from the prediction model is compared with label data (label data, true value, $y^{(i)}_{true}$), and RMSE (Root Mean Squared Error) is obtained as in Equation (7). We use this as a cost function and the optimizer learns the prediction model using Adam.

<Equation 7> RMSE (Root Mean Squared Error) cost function

$$Cost(W,b) = \sqrt{\frac{1}{n}\sum_{i=1}^{n}(y_{pred}^{(i)} - y_{true}^{(i)})^2}$$

($y_{pred}^{(i)}$ : predicted value, $y_{true}^{(i)}$ : true value)

There are variety of indices used. In addition to quantitative indicators at the present time, it also enhances the expression of the input data set by utilizing the past trends that have been frequently used in the financial market, and the auxiliary index expressing the current state. The number of auxiliary indicators is n, and the quantitative data is connected by using m. And sentiment data uses k data from related articles and data from sentiment analysis. In the case of financial indicators such as NASDAQ 100, social media and article data are selected as social media articles in the financial market as indicators. For individual companies, use the top k social media and articles for a specific date of the company.

[Figure 5] summarizes the overall structure of the proposed system. In the proposed system, financial market trends are predicted by using both auxiliary index, quantitative index and sentiment data. The reason for using multiple secondary indicators and quantitative data rather than one indicator is to understand the correlation between indicators and data. For example, when the RSI indicator is above 30, multiple indicators are used together to study the relationship between the secondary indicator and the quantitative indicator, as the volume increases.

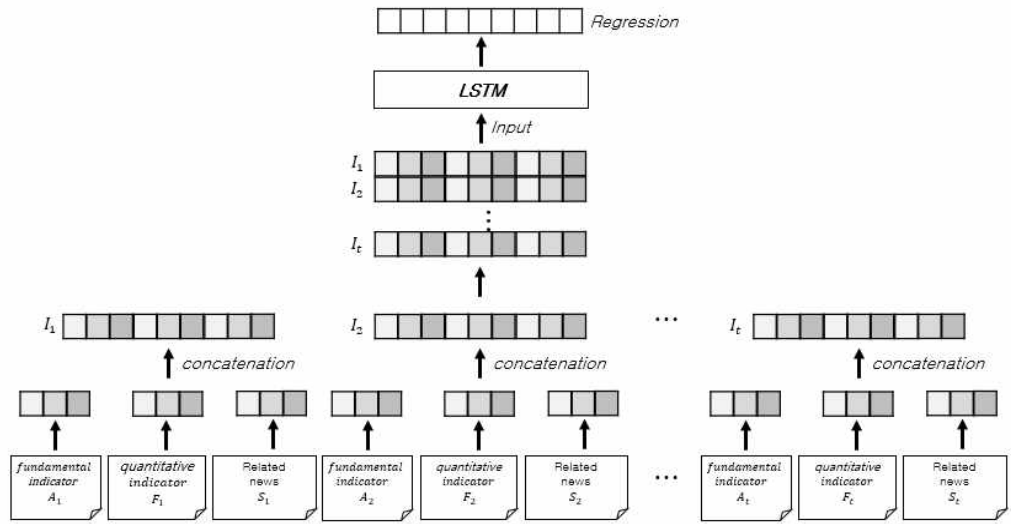

[Figure 5] Overall structure of analytical system for forecasting financial market trend

We make a total connection data by connecting each input data and use it as the input value of the LSTM to predict financial market trend through linear regression analysis. The concrete data used in the experiments of this paper are covered in Chapter 4 Experiments and Results.

## 4. Experiments and Results

This section describes various experiments and the results of the proposed prediction technique. Section 4.1 describes the Data Sets, Section 4.2 describes the Implementation, and Section 4.3 describes the Comparison of the Experimental Results.

### 4.1 Data Sets

In this paper, we use PBR (Price Book Ratio), PER (Price Earnings Ratio) and PSR (Price Sales Ratio) among subsidiary indicators related to corporate value for fundamental analysis. Enterprise fundamental and value analysis play a significant role in predicting actual commodity prices and financial market trends. However, since it is difficult to systematically include financial statements, announcements, related articles, and CEO's capabilities, we use indicators related to corporate values that are often used in financial markets.

Since the following figures can not be obtained when analyzing indices and financial indices other than individual companies, we use modified stock prices, trading volume, and Trend Deterministic Data (TDD) data. The TDD data is the difference between the previous day and the current stock price.

In the case of the commodity price-related data (adjusted price), since the size varies depending on the company or the market, the commodity price related data uses the normalized value using Equation (8). The normalized value $value_{c_n}^t$ is the value between [-1,1]. This $c_n$ is the firm's commodity price at time t, where $\max_{c_n}$ is the maximum value of the commodity price over the whole data period, and $\min_{c_n}$ is the minimum commodity price for the entire data period.

<Equation 8> Normalization process of data of commodity price

$$value_{c_n}^t = \frac{2*price_{c_n}^t - (\max_{c_n} + \min_{c_n})}{\max_{c_n} - \min_{c_n}}$$

<Table 1> Data formats for analyzing indices and financial indicators

| Adj. Price | Trading Vol. | TDD | RSI | CCI | MACD | Sentiment | Answer |
|---|---|---|---|---|---|---|---|
| 1728.339966 | 6610950000 | 66.448631 | 57.202 | 14.8892 | 6.880 | 0.802898 | 1814.790039 |
| 1814.790039 | 7999170000 | 86.450073 | 65.884 | 130.5910 | 27.080 | 0.812598 | 1803.47998 |
| 1803.47998 | 10466210000 | -11.310059 | 40.303 | -67.8739 | -3.290 | 0.701859 | 1875.380005 |
| 1875.380005 | 10514210000 | 71.900025 | 67.103 | 185.5059 | 163.930 | 0.608462 | 1864 |
| 1864 | 10541390000 | -11.380005 | 45.466 | -77.3871 | 5.650 | 0.901593 | 1902.880005 |

<Table 2> Data Formats for Individual Company Analysis

| PBR | PER | PSR | RSI | CCI | MACD | Sentiment | Answer |
|---|---|---|---|---|---|---|---|
| 1.06 | 38.62 | 2.10 | 61.542 | 120.2863 | 0.450 | 0.759621 | 1728.339966 |
| 1.09 | 38.72 | 2.15 | 60.741 | 86.0945 | 0.200 | 0.953152 | 1814.790039 |
| 1.04 | 38.65 | 2.08 | 57.745 | 60.2079 | 0.190 | 0.325489 | 1803.47998 |
| 1.05 | 38.75 | 2.09 | 52.559 | 122.3601 | -0.030 | 0.465923 | 1875.380005 |
| 1.11 | 38.81 | 2.16 | 48.174 | -77.7778 | 0.020 | 0.531658 | 1864 |

For technical analysis, we use indicators that are mainly used for technical analysis of stock prices. In addition to fundamental analysis data, this study also requires data for technical analysis. [Figure 6] is a supplementary indicator for the technical analysis used in Yakup and Jigar's study [29]. These studies used moving average line (MA), momentum stochastic index (K%, D%), RSI (Relative Strength Index), MACD (Moving Average Convergence Divergence), Larry William's R% and CCI (Commodity Channel Index). Among these data, we use RSI, CCI, and MACD data for technical analysis. The interval setting of the auxiliary index is based on 'weekly data'. The description of each indicator is covered in [29] and is not covered in detail in this paper.

| Name of indicators |
| --- |
| Simple $n(10here)$-day Moving Average |
| Weighted $n(10here)$-day Moving Average |
| Momentum |
| Stochastic $K\%$ |
| Stochastic $D\%$ |
| Relative Strength Index (RSI) |
| Moving Average Convergence Divergence (MACD) |
| Larry William's R% |
| A/D (Accumulation/Distribution) Oscillator |
| CCI (Commodity Channel Index) |

[Figure 6] Yakup, Jigar Research Supplementary Indices

<Table 3> Price, Trading Volume, Adjusted price, TDD

| Date | Mkt. Price | High Price | Low Price | Closed Price | Adj. P. | Trad. Vol. | TDD |
|---|---|---|---|---|---|---|---|
| 2010-06-28 | 1761.97998 | 1776.609985 | 1700.040039 | 1728.339966 | 1728.339966 | 6610950000 | |
| 2010-07-05 | 1752.97998 | 1815.23999 | 1719.199951 | 1814.790039 | 1814.790039 | 7999170000 | 86.450073 |
| 2010-07-12 | 1814.48999 | 1863.52002 | 1802.540039 | 1803.47998 | 1803.47998 | 10466210000 | -11.310059 |
| 2010-07-19 | 1807.98999 | 1875.380005 | 1784.550049 | 1875.380005 | 1875.380005 | 10514210000 | 71.900025 |
| 2010-07-26 | 1875.77002 | 1900.150024 | 1833.900024 | 1864 | 1864 | 10541390000 | -11.380005 |
| 2010-08-02 | 1886.609985 | 1911.01001 | 1873.439941 | 1902.880005 | 1902.880005 | 9623140000 | 38.880005 |
| 2010-08-09 | 1911.380005 | 1918.780029 | 1807.459961 | 1818.800049 | 1818.800049 | 9723350000 | -84.079956 |
| 2010-08-16 | 1806.969971 | 1862.369995 | 1801.73999 | 1825.75 | 1825.75 | 9012350000 | 6.949951 |
| 2010-08-23 | 1833.329956 | 1842.790039 | 1747.319946 | 1791.640015 | 1791.640015 | 9766090000 | -34.109985 |

As shown in Table 3, the price, trading volume, and business index data used in the experiment are the daily/weekly data of NASDAQ 100 from January 7, 2000 to September 12, 2017 provided by Yahoo Finance [21] daily/weekly data. The secured data include date, market price, high price, low price, close price, adjusted price, trading volume, and TDD.

The adjusted price is data that is more useful than other price data in data analysis because it guarantees continuity with stock price data that reflects dividend, allocation, new stock issue. In this paper, we use adjusted prices to increase the accuracy of data and use it as label data. In addition, price data is very meaningful data in addition to price data because the trading volume data generally precedes the stock price change and the trading volume increases.

Since learning data and test data are required to test the prediction results after learning the prediction model, the entire data must be divided at a certain rate. In this paper, the prediction model was studied using data from January 2000 to the end of 2015 at a ratio of 15:1, and the prediction results were measured using the data from January 2016 to February 2017 as test data .

We use enterprise tickers to identify companies, and bring the sentiment data analyzing industries, fundamental data, indicators for technology analysis (RSI and CCI, MACD), social media, and news. We crawled in Yahoo Finance using a library called python's BeautifulSoup. In this paper, we conduct experiments on individual company data in <Table 4>. We chose the representative companies of various industries and companies with large market cap.

Social media and article data for sentiment analysis are used by crawling top articles in the social and articles area provided by Yahoo Finance. To extract sentiment data from social media and article data, IBM Bluemix [20] was used. Using AlchemyLanguage from IBM Bluemix, extract data such as Entity Sentiment, Document Sentiment, etc. as shown in [Figure 7]. In this paper, we use affirmative / negative / neutral figures using the top 5 related social media and article data. The closer the value of sentiment data is to 1, the more positive the signal.

[Figure 7] IBM Bluemix sentiment analysis results

### 4.2 Implementation

The prediction system implementation language was Python. Python is often used in financial data analysis because it has the advantage of being able to handle all three data acquisition, logic analysis and fulfillment phases required for financial data analysis. Data acquisition in Python can use the Web Data Extraction library (Beautiful-Soup, Request, StringIO), and also supports connections to relational databases such as MysqlDB [30]. The logic analysis area also supports python libraries, numpy [31] and pandas [32], which deal with numbers in relational database format. Statsmodel [33], tensorflow [34], keras [35], and pytorch [36]. It also supports

Matplotlib [37] as a visualization tool. In addition, most securities firms' API (Application Programming Interface) have a variety of libraries (Xlsxwriters) that can be easily used with Python's win32com COM object and output Excel-formatted reports.

The machine learning algorithm library needed to implement the LSTM was used with keras and tensorflow libraries. The learning environment used Google Cloud AI [26] and the optimizer Adam [25]. Learning generation, learning rate, and model layer can be set as hyper parameters. (2000, 0.01, and 3 respectively in the experiment). We used RMSE (Root Mean Squared Error) as a cost function.

### 4.3 Comparison of Experimental Results

#### 4.3.1 Performance Comparison with Existing Financial Market Trend Forecasting Models

In this paper, we compare the performance of the financial market trend prediction method using the LSTM and the existing prediction model. The data used the same conditions. The comparison subjects are the study using ANN, SVM, and RNN.

In this experiment, we compared the performances according to the data types with the existing prediction models, compared the performances according to the financial market environment, and compared the performance according to the industry.

### 4.3.1.1 Performance Comparison with Existing Models by Data Type

When we compare the prediction performance according to the data type, the performance of the circular neural network is best when the daily data is used as shown in [Figure 8]. When the weekly data were used, the model using the LSTM specialized for long-term memory showed the best performance. In addition, the performance of the LSTM is significantly improved when the weekly data is used compared to the case of using the daily data. It can be seen that the LSTM is suitable for long data period compared to existing prediction models.

<Table 5> Performance comparison with existing model according to data type

| Type | ANN[22] | SVR[23] | RNN[24] | LSTM |
|---|---|---|---|---|
| Daily Data | 0.3398 | 0.3316 | 0.3015 | 0.3276 |
| Weekly Data | 0.3015 | 0.2863 | 0.2985 | 0.2516 |

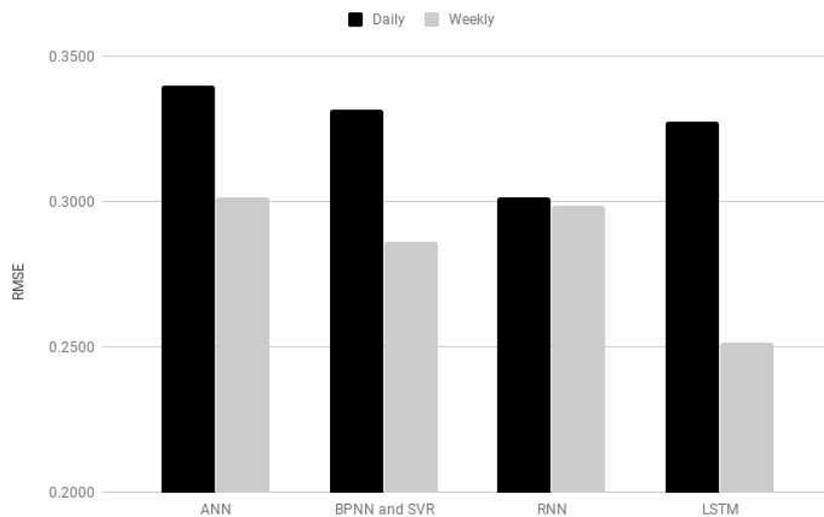

[Figure 8] Performance comparison with existing model according to data type

### 4.3.1.2 Performance Comparison with Existing Models by Financial Market Environment

[Figure 9] compares the performance of existing trend prediction models and trend prediction models of this paper in various financial market environments. The data section is made up of weekly data, and the experiment is divided into bull, bear, and flat trend sections (see Section 4.3.5). As a result, the technique using the LSTM shows the best performance in both bull, flat, and bear trends. All of the comparative examples show that the performance is better in the environment of the upward and downward trends than the environment of the flat trends. Therefore, the prediction model shows a relatively good performance in a market environment with large variation.

<Table 6> Performance comparison with existing model according to financial market environment

| Market Environment | ANN[22] | BPNN, SVR[23] | RNN[24] | LSTM |
|---|---|---|---|---|
| Bull/Weekly | 0.3206 | 0.3105 | 0.3069 | 0.2596 |
| Flat/Weekly | 0.3695 | 0.3659 | 0.3326 | 0.2910 |
| Bear/Weekly | 0.3295 | 0.3462 | 0.3215 | 0.2411 |

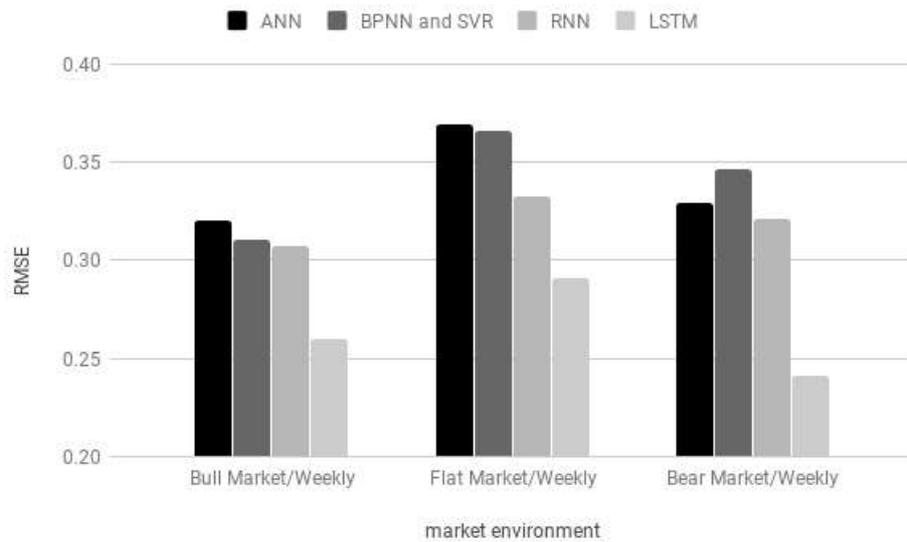

[Figure 9] Performance comparison with existing model according to financial market environment

#### 4.3.1.3 Performance Comparison with Existing Models by Industry

[Figure 10] compares the performance of existing models with those of existing models in financial markets. Companies with IT and semiconductor, cosmetics, pharmaceuticals, automobiles, and chemicals are ranked in the order of the largest number of companies classified by industry. Individual company data is selected by firms with high market capitalization, which represents the overall financial market trend (see Appendix A). The forecasted performance is measured using the top 5 companies with high market capitalization by industry, and the average of the total performance is obtained. The time interval of the data is also the weekly data. The results show that the technique using the LSTM is the best in all industries. Specifically, it can be seen that both the existing techniques and the techniques of the present invention show the best performance in the pharmaceutical industry. This also shows that both the existing techniques and the techniques of this paper work well in environments or industries where trends within the period are clear.

<Table 7> Performance comparison with existing model according to industry

| Sectors | ANN[22] | BPNN, SVR[23] | RNN[24] | LSTM |
|---|---|---|---|---|
| IT&Semi-con. | 0.3166 | 0.2856 | 0.2983 | 0.2715 |
| Cosmetics | 0.3148 | 0.2911 | 0.3015 | 0.2659 |
| Healthcare | 0.3026 | 0.2806 | 0.2796 | 0.2496 |
| Automobile | 0.3085 | 0.2929 | 0.2813 | 0.2548 |
| Chemistry | 0.3265 | 0.2866 | 0.2948 | 0.2658 |

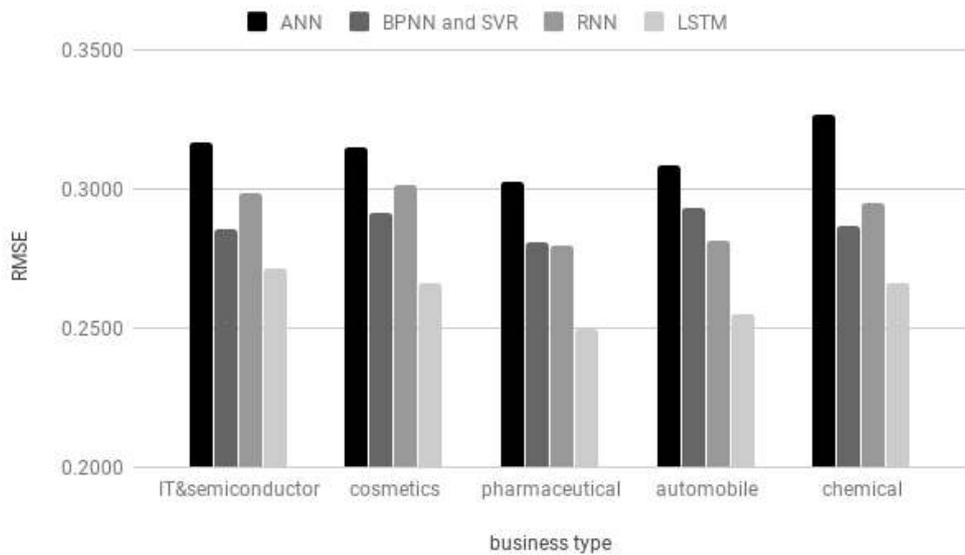

[Figure 10] Performance comparison with existing model according to industry

### 4.3.2 Performance Comparison by Interval of Data

In order to evaluate the predictive performance of the proposed model

- 24 -

according to the data interval, other variables except the interval of the data are fixed and the data use the NASDAQ 100 financial index. [Figure 11] shows the comparison of prediction performance according to the interval of the data interval. The horizontal axis shows the type of data according to the interval, and the vertical axis shows the RMSE. As a result, the data section was experimented with weekly data (NAQ_100_W). As a result, the RMSE of the training data was 0.2783 and the RMSE of the test data was 0.2825. Experimental results of data section with daily data (NAQ_100_D) show that RMSE of training data is 0.3750 and RMSE of test data is 0.3792.

Overall, the performance of week intervals was better than that of days. It can be seen that the data with longer time intervals is more suitable for the financial market prediction model using the LSTM.

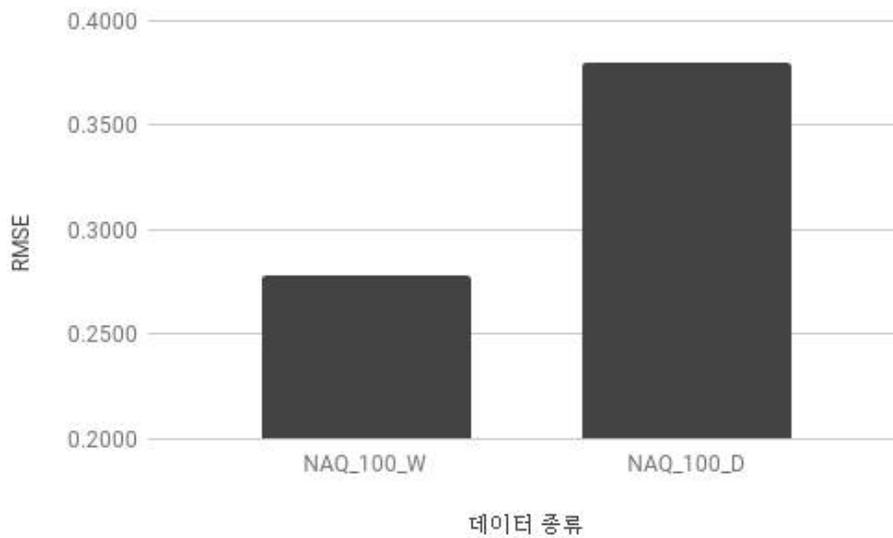

[Figure 11] NASDAQ 100 daily and weekly data performance comparison

### 4.3.3 Performance Comparison According to Various Financial Market Environments

Experiments are conducted using NASDAQ 100 data to determine if the predictive models show performance differences in predicting trends in financial markets in diverse environments. There are four types of long-term fluctuations: bull, bear, flat, and volatility. In this paper, three kinds of bull, bear, and flat strength are divided into sections and the experiment is conducted. The uptrend refers to the steady increase that keeps the bottom low, while the downward steepness refers to the steady decrease that keeps the peak. The flat period refers to a period in which the bottom and top move up and down and move horizontally. In this paper, we divide the tax base and set the same period (2 years) as shown in [Figure 16], so that we can confirm the performance difference according to the financial market environment, not the difference in performance due to the amount of data.

Data were divided into training data and test data at a ratio of 15:1 in order to estimate the performance of forecasting models for each financial market environment. [Figure 17] compares the performance of the models according to the financial market environment. It can be seen that the trend of the financial market, which is lagging from September 2004 to August 2006, shows lower prediction performance than other environments. On the other hand, the trend of the financial market, which has been declining from February 2000 to January 2002, shows better performance than other environments. In addition, the financial market trend that has been rising from August 2013 to July 2015 shows relatively better performance than the flat period. Although a simple comparison between the segments may be difficult because of various times and many factors, it can be seen that the forecasting model of this paper works well in the financial market environment where a large rise or fall change generally occurs.

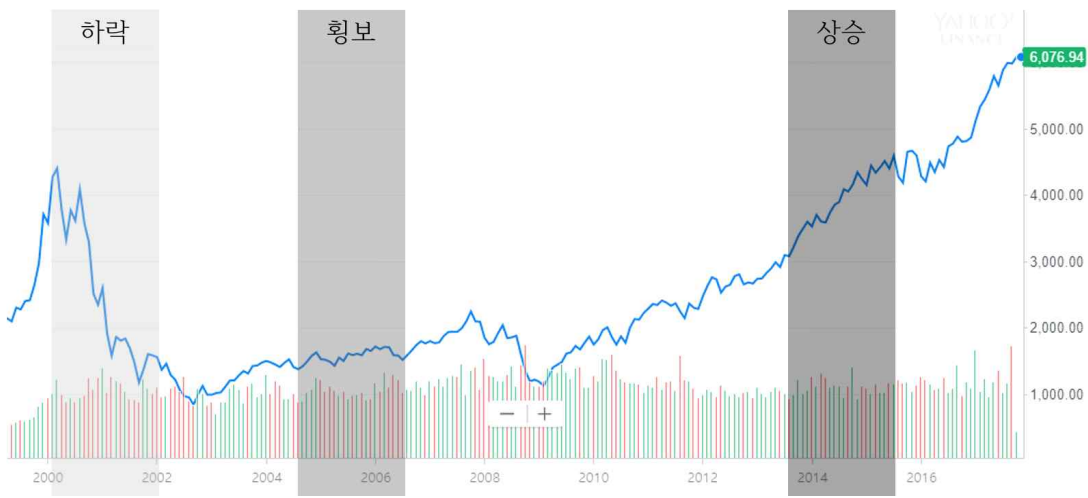
[Figure 12] From 2000 to 2017, NASDAQ 100's bull, flat, and bear.

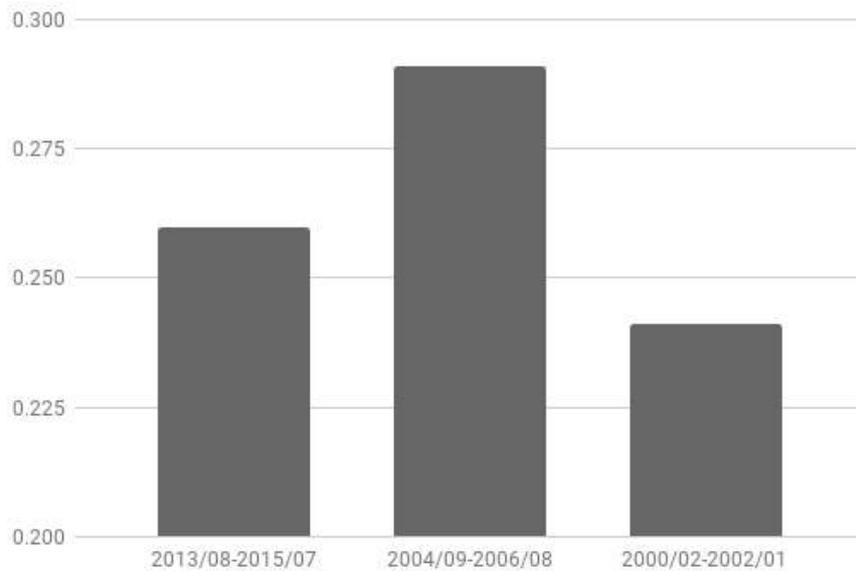

[Figure 13] Comparison of model performance according to financial market environment

### 4.3.4 Performance Comparison According to Sentiment Analysis Inclusion

In order to verify the validity of the sentiment data, we compared the case where the sentiment data is included in the input data and the case where the sentiment data is not included. There were no significant differences in Support Vector Regression between the two cases. However, in the case of RNN and LSTM, sentiment data included in the input data showed better performance. This is because the source of the sentiment data is news and SNS data by date, and therefore the sentiment data also has time-series characteristics.

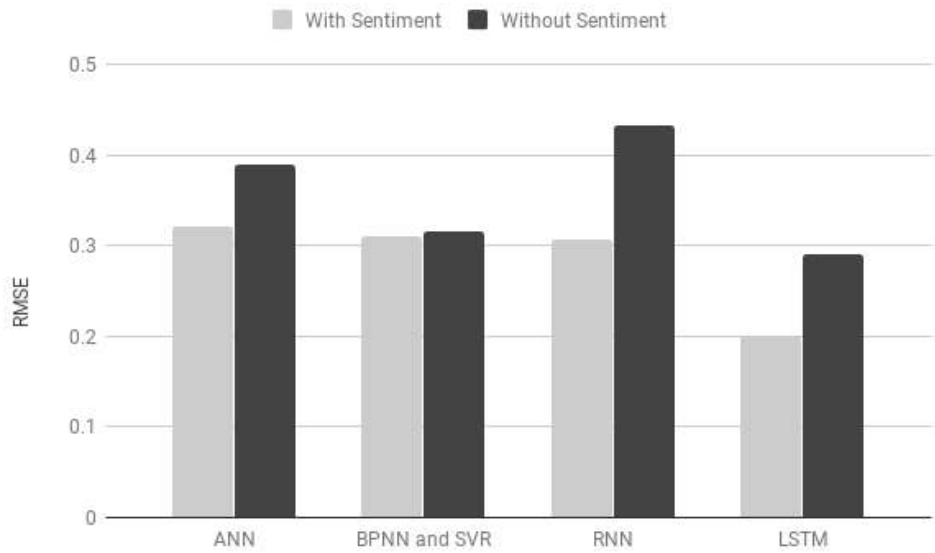

[Figure 14] Performance Comparison According to Sentiment Analysis Inclusion

### 4.3.5 Analysis of Forget Gate Value

In order to verify that LSTM model works well, we compared the forget gate value according to the time window size. Experimental results show that the forget gate value increases monotonously according to the window size. That is, the forget gate value is related to the window size. As the window size is shorter, the forget gate value is smaller because it is a prediction that does not require long historical data. On the contrary, as the window size increases, the long historical data is required. The larger the forget gate value, the more long historical data is retained.

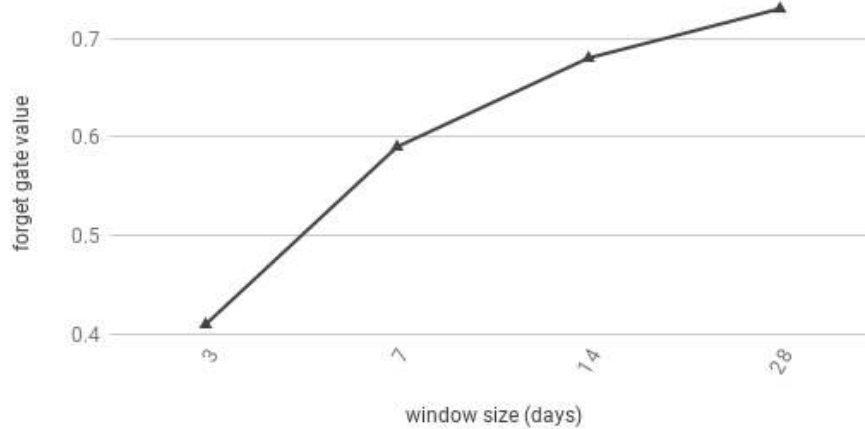

[Figure 15] Analysis of Forget Gate Value

## 5. Conclusion and Suggestion

In this paper, we propose a financial market forecasting method using LSTM, and apply the fundamental analysis, technical analysis, and sentiment analysis data used in predicting commodity prices and financial market trends to input values. Respectively. In addition, the performance of the proposed method is verified by comparing with various experiments and existing prediction models. It is likely that the data will be further improved by further modeling the data or by optimizing the prediction techniques more structurally.

However, since the actual financial market trends are different depending on the age and various factors are influential, it is not possible to predict the financial market trend because it shows meaningful results using only historical data. In addition, we can not overstate this forecasting system because historical data does not guarantee future data, and analysis of financial market based on strategy is based on the assumption that "the past will recur". Since the technique using the neural network basically has a 'black box' structure in which the result is not known through the process, there is a certain limit to be used in decision making in the field. Therefore, in the field, machine learning techniques are often used for analysis of alternative data such as unstructured data processing, natural language processing, social media analysis, voice data analysis, satellite image analysis using image processing (eg. parking lot).

This paper does not deal with what kind of data is appropriate to use in financial market trend forecasting, but it proposes trend forecasting method of financial market using LSTM. Considering the fact that it is difficult to achieve performance of more than 80% in the prediction task of financial market using machine learning in the existing study, the trend forecasting model of the financial market using the LSTM proposed in this paper is based on the existing forecasting models. The results are satisfactory. In addition, we show that financial market data with time-series data works well with techniques using LSTM, compared with other schemes using other structures.

In addition, the initialization method of the prediction model and the experiment of how to solve the over sum problem (dropout) in the financial data, the correlation coefficient analysis between the index and individual companies, and the performance of the existing prediction model from various analysis will be helpful in conducting research on future financial market trend forecasting techniques.

Furthermore, It is very meaningful that we proposed model that can predict the financial market trend through various combinations of fundamental analysis data, technical analysis data, and sentiment analysis data, and compared with the existing complex mathematics and statistical financial models.